\documentclass[prb,aps,twocolumn,floats,showpacs,amsmath,amssymb]{revtex4-1}
\usepackage{bm,graphicx}
\usepackage{amsmath}
\newcommand{\be}{\begin{equation}}
\newcommand{\ee}{\end{equation}}
\newcommand{\bea}{\begin{eqnarray}}
\newcommand{\eea}{\end{eqnarray}}

\newcommand{\phrb}[1]{Phys.~Rev.~B {\bf #1}}

\newcommand{\bib}{\bibitem}

\usepackage{graphicx}
\usepackage{color}
\usepackage{epsfig}
\usepackage{epstopdf}
\usepackage{amsmath,amssymb}
\usepackage{hyperref}

\begin{document}

\title{Exciton condensation in an extended spinless Falicov Kimball model in the presence of orbital magnetic fields}
\author{S. Pradhan$^1$}
\email[E-mail: ]{spradhan@phy.iitkgp.ernet.in }
\author{ A. Taraphder$^{1,2}$ }
\affiliation{ $^1$Department of Physics, Indian Institute of Technology Kharagpur, Kharagpur - 721302, India \\ $^2$Center for Theoretical Studies, Indian Institute of Technology Kharagpur, Kharagpur - 721302, India}

\begin{abstract}
An extended, spinless Falicov-Kimball model in the presence of perpendicular magnetic field is investigated employing Hartree-Fock self-consistent mean-field theory in two dimensions. In the presence of an orbital field the hybridization-dependence of the excitonic average ${ \Delta =<{{d_i}^\dagger} {f_i}>}$ is modified. The exciton responses in subtle different ways for different chosen values of the magnetic flux consistent with Hofstadter's well-known spectrum. The excitonic average is suppressed by the application of magnetic field. We further examine the effect of Coulomb interaction and $f$-electron hopping on the condensation of exciton for some rational values of the applied magnetic field. The interband Coulomb interaction enhances the $\Delta$ exponentially, while a non-zero $f$-electron hopping reduces it. A strong commensurability effect of the magnetic flux on the behaviour of the excitons is observed.

\end{abstract}
\date{\today}
\maketitle

\section{Introduction}
The Falicov-Kimball model is perhaps the simplest model to study correlation in fermionic systems on a lattice. It involves a conduction $d$-band, a localized $f$-electron state and an on-site Coulomb interaction ${U}$ between the $d$ and $f$-electrons. Since its introduction\cite{FKM1} in 1959, to describe valence or metal-insulator transition in some transition metal oxides, the model has been used successfully in describing various many-body effects~\cite{Freericks} like metal-insulator transition, \cite{FKM2} mixed-valence phenomena,\cite{FKM3} the formation of ionic crystals \cite{Gruber,Lemanski} and orbital \cite{Yadav1} and charge-density waves (CDW). \cite{Kennedy} It was found  \cite{Brandt, Kennedy, Lieb} that on a bipartite lattice at half-filling ${({n_d} = {n_f} = 0.5)}$,\, ${f}$-electrons occupy sites of one sublattice only, the well-known checkerboard phase. For ${D\ge 2}$ dimension, the chessboard charge pattern exists below a critical temperature, $T_{CDW}$, above which a disordered phase is obtained. \cite{Freericks} Further, using dynamical mean field theory (DMFT),\cite{DMFT} exact in dimension ${D= \infty}$, Brandt and Mielsch confirmed the existence of inhomogeneous CDW phase.\cite{Mielsch} An extended version of Falicov-Kimball model is also in use to account for the homogeneous mixed valence problem. It has been used since last two decades to account unconventional ferroelectricity \cite{Leder,Portengen1,Portengen2} in mixed-valent compounds.

Customarily, ferroelectricity appears due to the structural phase transition. However, it is also possible that there is a nonvanishing $d-f$ coherence in a system leading to a hybridization term of purely electronic origin. This coherence could give rise to electric polarization due to Bose-Einstein condensation of ${d-f}$ excitons when the two bands differ by odd parity. Portengen and co-authors studied an extended FKM with a ${k}$-dependent hybridization term in Hartree-Fock approximation following Leder's work. \cite{Leder} They found that the Coulomb interaction $U$ between itinerant $d$-electron and localized $f$-electron gives rise to a nonvanishing ${d-f}$ coherence $<{{d_i}^\dagger}{ f_i}>$ even in the limit of vanishing hybridization $V\to 0$ in the presence of a putative homogeneous ground state solution of the HF approximation. In the weak-coupling mean-field theory, the formation of an order parameter and the condensation thereof are concomitant. Quite interestingly, this condensation of the excitonic order parameter, they pointed out, signifies a ``spontaneous'' polarization in the system when the parities of the electron-hole partners in the condensate differ by one.

Later, Czycholl \cite{Czycholl} showed that an imhomogeneous ground state, as obtains in FKM on a square lattice at half-filling, leads to $<{{{d_i}^\dagger}{f_i}> \to 0}$ as ${V \to 0}$. Therefore, the FKM in the half-filled limit does not admit a ``spontaneous" symmetry breaking, consistent (but not contradictory at $T=0$, where Elitzur's theorem does not forbid an order) with the local $U(1)$ symmetry in the $f$-band at $V=0$. For a small non-zero hybridization $V$, the inhomogeneous (CDW) phase is stable, and the order parameter is finite. This CDW phase, though, melts beyond a critical hybridization strength. Similar conclusions were reached in a triangular lattice as well~\cite{Yadav2}. For a one-dimensional extended FKM, using exact-diagonalization and DMRG techniques, Farka\u sovsk\'y has ruled out the possibility of spontaneous excitonic averages at zero temperature \cite{Farkasovsky12}, which is expected in one dimension. Employing the same numerical technique, Farka\u sovsk\'y  investigated the effects of local and non-local \cite{Farkasovsky3, Farkasovsky4} hybridization on valence transitions. Zlati\'c et al.\cite{Zlatic} confirmed that the static excitonic susceptibility diverges at $T=0$ in the ordinary FKM $(V=0)$, from an exact solution of the model in infinite dimension. In dimensions $D>1$, a finite $f$-electron bandwidth breaks the local $U(1)$ symmetry and induce a non-zero polarization even in the absence of ${d-f}$ hybridization as expected from symmetry grounds. This is easily shown \cite{Batista1} by mapping an extended FKM ($V=0$ but $t_f$ finite) on to a Hubbard model with asymmetric hopping ($t_{\uparrow} \ne t_{\downarrow}$) and thence to an effective anisotropic $XXZ,\, s=1/2$ spin model with a ``field" along z-direction in the large $U$ limit. The intermediate coupling regime was treated with constrained path Monte-Carlo (CPMC) technique. A nonlocal hybridization in an extended FKM stabilizes excitonic averages with the inclusion of $f$-electron hopping.\cite{Batista1, Batista2, Sarasua} 

Ferroelectric materials have sustained a significant attention in condensed matter physics and have been an important component of modern applications and technology. In conventional ferroelectrics, ferroelectricity is connected with distortions of the lattice. In multiferroic materials this can be induced by magnetic field which has spawned a considerable interest owing to the tunability of the ferroelectric order through magnetic fields. The unconventional ferroelectricity arising out of $d-f$ coherence is, on the other hand, purely of electronic origin and offers an additional route to tuning optical properties with magnetic field. \cite{Portengen2}

A moot question is therefore what happens to such an excitonic condensate in the presence of a strong orbital field. This kind of gauge-field can be experimentally realized by using ultracold particles (fermions and bosons) on optical lattices.\cite{Dalibard} Moreover, there are recent proposals for the realization of FKM in optical lattices with mixtures of light atoms in the correlated disordered environment formed by heavy atoms.\cite{Ates, Ziegler} Therefore, it is quite pertinent to appraise an extended FKM in optical lattices in the presence of an artificial gauge field produced in the same lattice. In a very strong magnetic field, the two Zeeman-split bands are well separated in energy and at low filling, only the lower band is relevant, effectively reducing it to a spinless problem. The field then couples to the orbital degrees only via the canonical transformation. We implement a self-consistent mean-field calculation using exact diagonalization to study the effect of a perpendicular magnetic field. We first examine the case without a magnetic field and then study the effect of orbital field on the exciton condensation. Results for both commensurate and incommensurate magnetic fluxes are obtained and discussed.

\section{Model}
We consider an extended FKM for spinless fermions on a square lattice represented by the Hamiltonian  
\begin{equation}
\begin{array}{l}
H =  - {\sum\limits_{ < ij > } {({t_{ij}}{d_i}} ^\dag }{d_j} + h.c) + \\
U\sum\limits_i {{d_i}^\dag } {d_i}{f_i}^\dag {f_i}  + {E_f}{\sum\limits_i {{f_i}} ^\dag }{f_i} \\+ {\sum\limits_i {V({d_i}} ^\dag }{f_i} + h.c)
\end{array}
\end{equation}

\noindent where ${<i,j>}$ are nearest-neighbour site indices on a square lattice (lattice parameter $a=1$ is chosen as length unit), $d_{i} \, (f_{i})$ are itinerant (localized) electron annihilation operators at site $i$. The first term represents the kinetic energy of $d$-electrons while the second term represents on-site Coulomb interaction between $d$-electrons of density ${n_d} = \frac{1}{N}\sum\limits_i {{d_i}^\dagger} {d_i}$ and the $f$-electrons with density ${n_f} = \frac{1}{N}\sum\limits_i {{f_i}^\dagger} {f_i}$);\, ${N}$ being the number of sites. The third term  ${E_f}$ represents the non-dispersive $f$-electron energy level. The fourth term stands for the hybridization between ${d}$ and ${f}$ electrons. We consider a model system with a ${d}$ band and ${f}$ level arising from ${d}$ and ${f}$ orbitals on every site. We set the hopping integral $t$ to be $1$ throughout the calculation and all other energy parameters are defined in units of $t$. In the absence of the hybridization term, the Hamiltonian commutes with $\hat{n}_{f,i}$, in which case $n_{f,i}$ is a good quantum number. In this case, the Hamiltonian is exactly solvable in infinite dimension. By annealing over $f$-electron distributions in the lattice, it can be `solved' numerically as well. The hybridization term $(V)$ removes the local $U(1)$ symmetry associated with the conservation of $n_{f,i}$ and the Hamiltonian is no longer exactly solvable, albeit in the above sense. We, therefore, take recourse to the usual Hartree-Fock self-consistent mean-field approximation to obtain the excitonic order parameter ${\Delta =<{{d_i}^\dagger}{f_i}>}$. 

In presence of a magnetic field, the spinless, mobile fermions `see' the field via the usual canonically conjugate momentum. On a lattice problem as at hand, the nearest-neighbour hopping term is therefore modified by a Peierls phase\cite{Peierls} factor. With the choice of Landau gauge $\vec{A}(r)={B(0,ma,0)}$ for a uniform magnetic field ${B}$ perpendicular to the plane of the lattice, the hopping integral $t_{ij}$ remains unchanged along the $x$-direction while $t_{ij}= -t\exp({\pm ie/\hbar\int_j^i{A(\vec{r})d\vec{r}}})$=$ {-t\exp ( \pm 2\pi im\frac{\phi }{{{\phi _0}}})}$ = $- t\exp ( \pm 2\pi im\frac{p }{{{q}}})$ along the $y$-direction. Here ${\phi}$=$Ba^{2}$ is the number of flux quanta per plaquette of a square lattice. This represents the gain of phase by an electron hopping round a closed path along the plaquette. Throughout our calculation, we consider only rational magnetic flux, i.e. $\frac{\phi }{{{\phi _0}}}= \frac{p}{q} = {\alpha}$ with ${p},\, {q}$ co-prime integers and $\phi_0$ is the Dirac flux quantum. Under the application of lattice translational operator, which moves ${q}$ number of lattice points along x-direction, the Hamiltonian remains invariant. 

Once the magnetic field is switched on, lattice periodicity is lost along $x$-direction. Due to the Peierls phases, Hamiltonian is not invariant to all lattice translations, only to those in the magnetic translation group. \cite{Zak} This group is associated with a magnetic unit cell $q$ times larger than the original unit cell, so as to enclose an integer flux $p{\phi_0}$. However, since it is only a phase, repeating at every ${2\pi}$, periodicity is still retained, albeit with a different value. Therefore, to accommodate a magnetic flux $B$=$\frac{2\pi}{N}$, the resultant magnetic supercell will be a strip of length $N$.~\cite{polaron} We chose a typical supercell of size ${24 \times 24}$ throughout the calculation. For the convergence criterion, we set the difference between the values of the order parameters for two consecutive iterations to within ${10^{-4}\%}$. The energy spectrum, in the non-interacting limit, is well-known and shows interesting structure: a self-similar structure in which the widths and gaps depend critically on the values of the magnetic flux.~\cite{Hofstadter} In the present work, we work with rational magnetic fields: in the non-interacting limit an irrational field induces a Cantor set.


\section{Results and Discussions}


\begin{figure}[h]
\includegraphics[trim={3cm 0cm 0cm 0}, clip, scale=0.32]{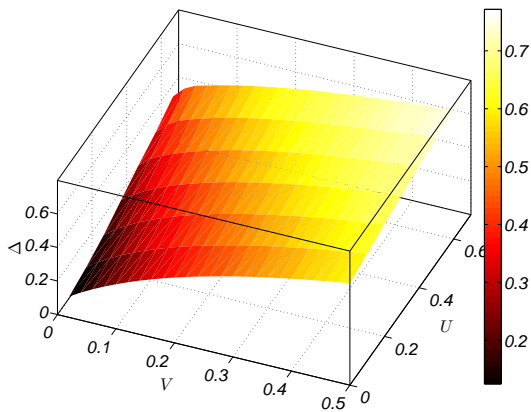}
\caption{\label{f1} (Color online)
The hybridization-dependence of excitonic order parameter ${\Delta = <{{d_i}^\dagger}{f_i}>}$ calculated for different values of ${U}$ in the zero-field limit for the symmetric case, i.e., ${E_f = 0}$. }
\end{figure}

 \begin{figure*}[h]
  \center
\includegraphics[trim={0cm 0cm 0cm 0cm}, clip, scale=0.55]{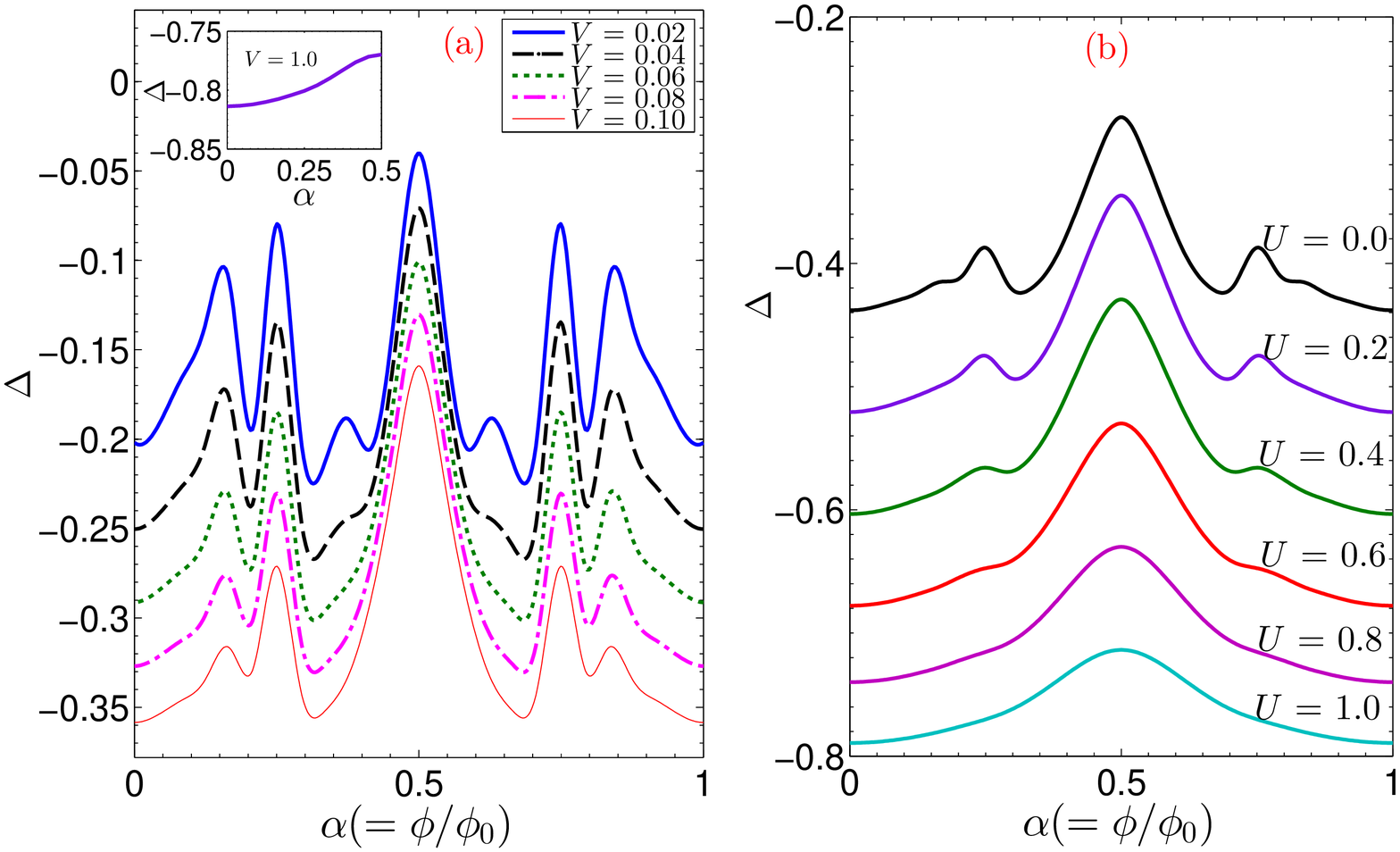}
\caption{\label{f2} (Color online)
(a) The variation of excitonic order parameter ${\Delta = <{{d_i}^\dag}{f_i}>}$ with magnetic field ${\alpha}$ at fixed ${U = 0.2}$ for several ${V}$. The inset shows the variation for a high $V$-value. (b)The same for a set of different ${U}$, with a fixed ${V = 0.20}$. Both are for the symmetric case, i.e., ${E_f = 0}$.}
\end{figure*}  

To check our numerical procedure we start with the extended FKM ${(H = H_0 + H_v)}$ in the symmetric case ($E_f = 0$ and $n_d = n_f=0.5$) without the transverse field. We study the effect of Coulomb interaction on the stability of excitons. The calculation shows there is no finite excitonic average in the ${V\to 0}$ limit as expected on symmetry grounds.~\cite{Czycholl,Yadav2} As we increase hybridization between $(d)$ and $(f)$ electrons for a fixed Coulomb correlation, we find an enhancement in the excitonic order parameter. Moreover, to see the effect of Coulomb interaction, we have chosen three different values of ${U}$. It is evident from Fig.1. that for finite ${V}$, there is a nonvanishing ${\Delta}$ that is strongly enhanced as ${U}$ increases. The effect of ${U}$ is stronger, smaller the value of ${V}$: the order parameter is exponentially enhanced as expected. These results are in complete agreement with previous results in a wide parameter regime.~\cite{Czycholl,Yadav2}

\begin{figure*}[t]
\includegraphics[trim={0cm 0cm 0cm 0}, clip, scale=0.7]{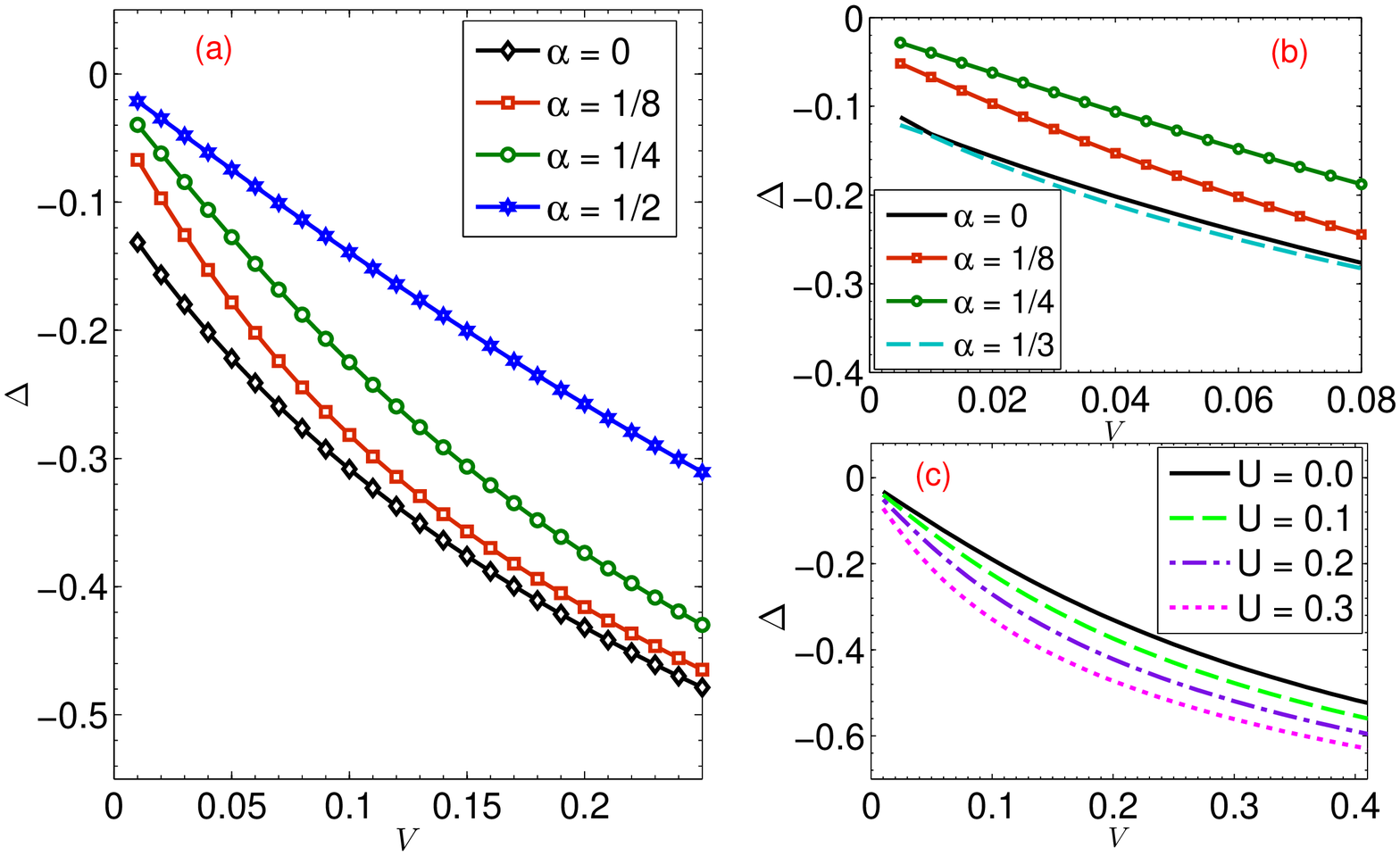}
\caption{\label{f3} (Color online)
 The variation of excitonic average with hybridization for different ${\alpha}$ values at ${U = 0.20}$. The dashed line in (b) shows `enhancement' of exciton average for ${\alpha = 1/3}$ case in the small $V$ limit., (c) shows the variation of ${\Delta}$ for different ${U}$ values for a fixed ${V ( = 0.2)}$ in presence of ${1/4}$ magnetic flux per plaquette.} 
\end{figure*}

The effect of orbital magnetic field on the evolution of excitons is a question we address next. With the inclusion of special flux values ${\alpha}$ per plaquette in a square lattice, the hopping term is modified by a phase and it affects the exciton formation via interference and localization effects, as seen elsewhere as well.~\cite{polaron,Bottger} As we see from Fig.2, the excitonic average is suppressed with the application of field. As it is clear from Fig.2(a), a prominent peak at ${\alpha = 0.25}$ and a dip at ${\alpha = 1/3}$ can be found for $U = 0$ case. There is a large peak at ${\alpha = 0.5}$; signifying the maximum value at which ${\Delta}$ is minimum and this is the case for all the values of $U$ studied. These undulations become lesser with larger $U$ and $V$ (see inset to Fig. 2a). The low $U$ and low $V$ regimes more or less follow the non-interacting physics and display a typical Hofstadter characteristics. As the Hamiltonian is symmetric with respect to ${\phi}$ and ${1 \pm\phi}$, Fig.2 is symmetric with respect to ${\alpha} = 0.5$ The variation of ${\Delta}$ with $V$ more or less follows the same pattern as in the absence of magnetic field (see Fig.3(a)). However, for a fixed ${U}$, the excitonic average reduces with increase in magnetic filed. It is also possible to get an enhancement in the excitonic average in a region where hybridization and correlation effects are quite weak (see  Fig.3(b)). 

\begin{figure*}[h]
\includegraphics[trim={0cm 0cm 0cm 0}, clip, scale=0.5]{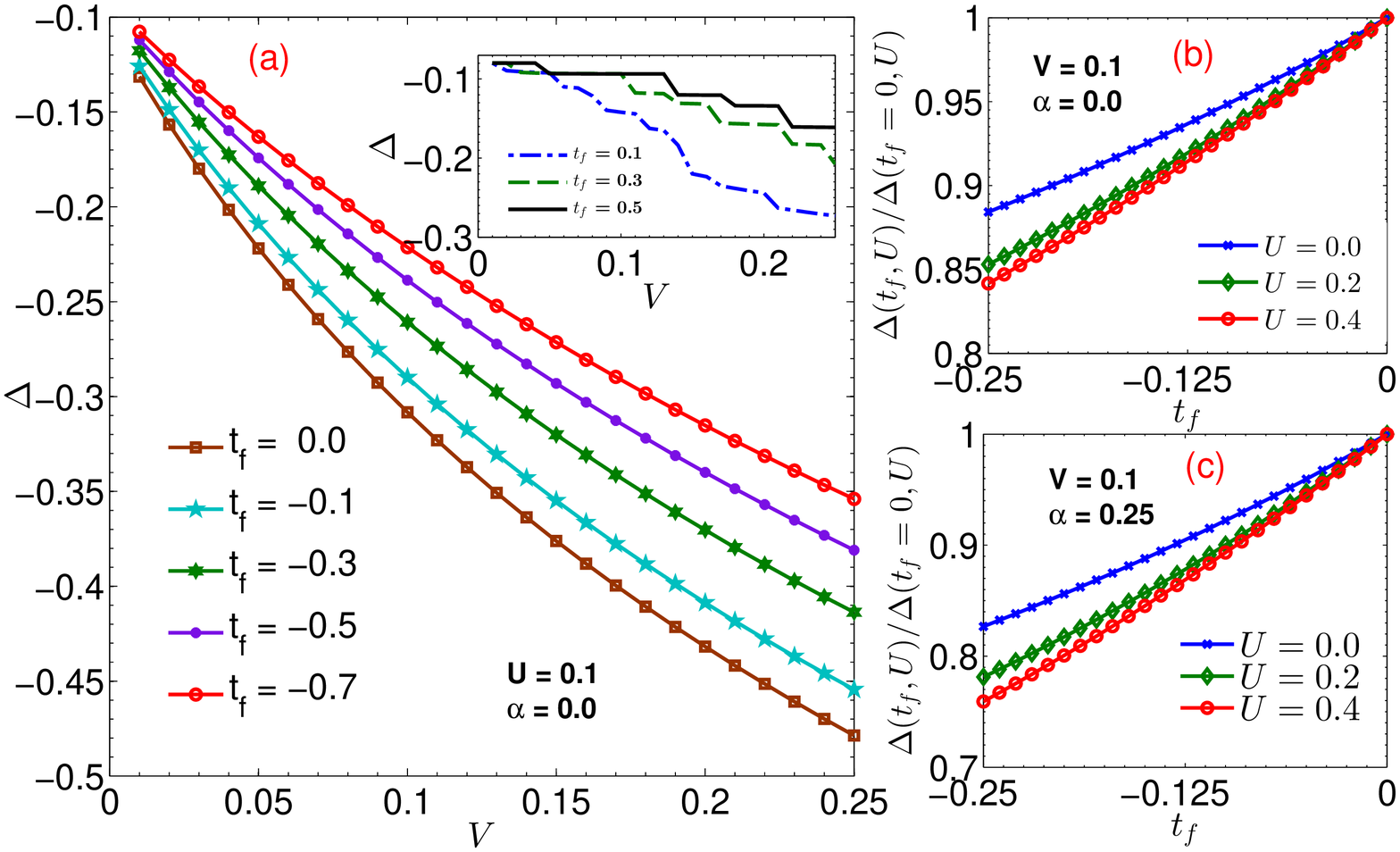}
\caption{\label{f4} (Color online)
(a)The variation of excitonic order parameter for different choices of negative ${t_f}$ values at a fixed ${U}$(= 0.1). The inset shows variation of excitonic average with positive ${t_f}$. Normalized value of order parameter as a function of $t_f$ for (b) $\alpha = 0.0$ and (c) $\alpha = 0.25$.}
\end{figure*}

So far, the $f$-electrons are kept localized without hopping. We now allow delocalization of the $f$-electrons with a dispersive term ${\sum\limits_{ < ij > } {({-t_{ij}}{f_i}} ^\dag }{f_j} + h.c)$ in the Hamiltonian. It is known that the $f$-electron hopping itself can give rise to non-zero ferroelectricity without the presence of explicit hybridization in the Hamiltonian in $D\ge 1$ dimension.~\cite{Batista1} We therefore examine the effect of $f$-electron hopping on the excitonic average. The magnitude as well as sign of $f$-electron hopping integral $t_f$ play important roles in the formation of an exciton. The value of excitonic order parameter is found to decrease with an increase in $t_f$; this is an effect of the enhanced kinetic energy of the $f$-electrons destabilizing the local excitonic order parameter. The same problem is then studied in the presence of an external magnetic field $\alpha = 0.25$. Fig.4(c) shows that the excitonic condensation is further suppressed with $\alpha$, increasing the kinetic energy of the $f$-electrons. As seen from Fig.4, for a fixed value of ${U}$, ${\Delta}$ drops with an increase in $t_f$. Clearly, this is expected as the $f$-electron occupancy at a site is lower now as compared to the case when it was occupied by a localized $f$-electron.
  
\begin{figure*}[t]
\includegraphics[trim={0 0cm 0cm 0cm}, clip, scale=0.4]{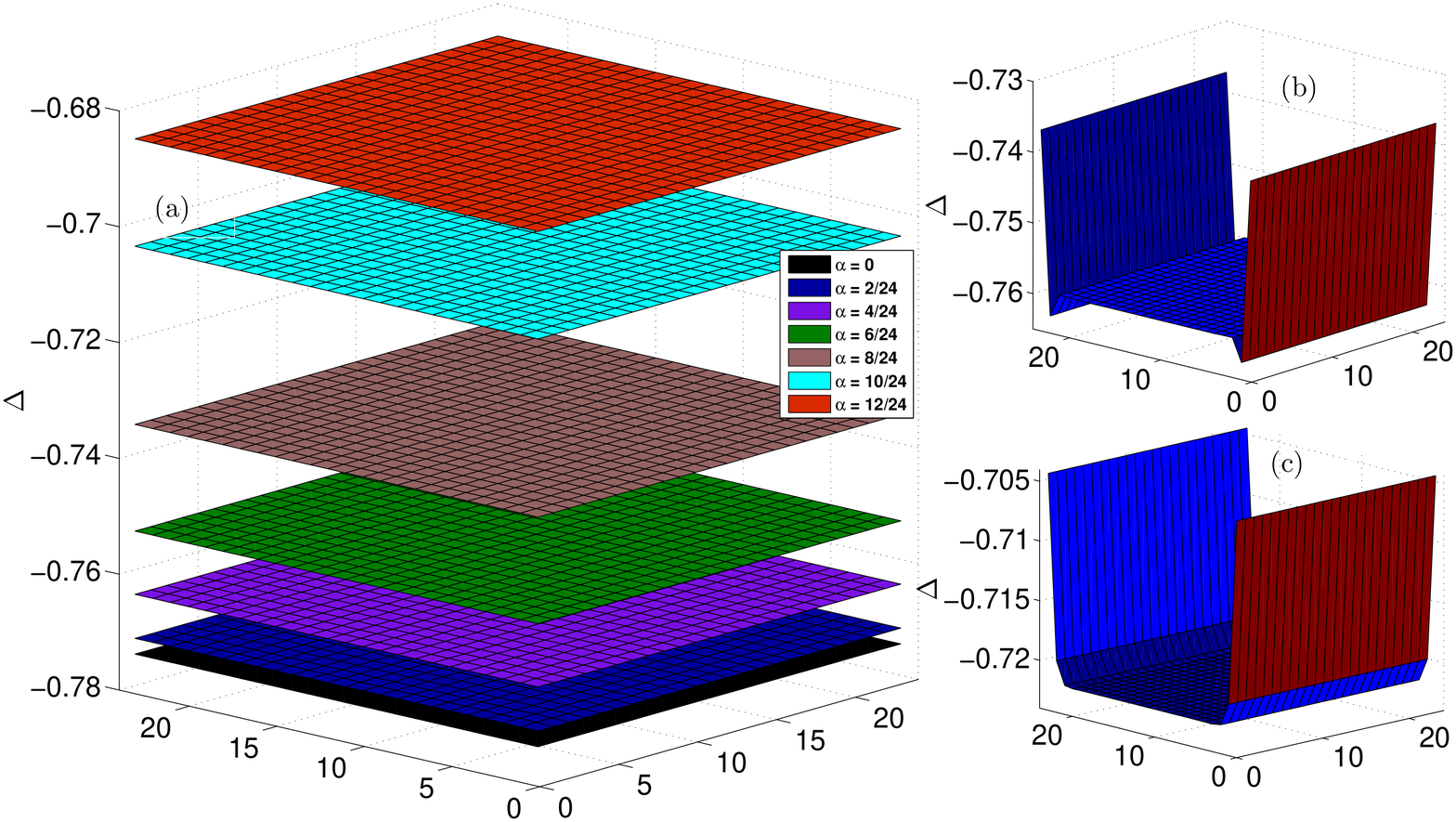}
\caption{\label{f5} (Color online)
The order parameter on a 2D lattice plane for (a) commensurate magnetic flux values, (b) ${\alpha = 0.1}$, (c) ${\alpha = 0.37 }$. With chosen $U =1.0$, ${t_f = 0.0}$ and $V = 0.2$.}
\end{figure*}

In the presence of the magnetic field, there is serious commensurability effect on the excitonic order parameter. With chosen values of $d-f$ correlation and hybridization terms, the system is essentially in the homogeneous ground state. As seen from Fig.5, for some magnetic flux values, the magnetic unit cell is commensurate with the lattice size (in the present study on a {24 $\times$ 24} lattice, ${1/24, 2/24, 3/24.. 12/24}$ and so on represent commensurate fluxes) and the excitonic average is uniform throughout the lattice. With increase in the commensurate flux values, magnitude of ${\Delta}$ decreases and the order parameter is uniform. On the other hand, the order parameter varies in a quite different way when the magnetic flux is incommensurate. In this case, the order parameter exhibits a one-dimensional modulation and the modulation length changes with the flux (Fig.5(b), (c)).

\section{Summary and Conclusion}
In the foregoing we have studied excitons in a transverse magnetic field numerically from an extended, spinless 2D lattice model, namely, the Falicov Kimball model. The FKM in presence of a hybridization term generally supports quantum mixed valence and an excitonic condensation of particle-hole bound states. In the presence of an orbital field, the itinerant electrons pick up a phase while the localized ones do not, and the excitonic condensate is expected to be affected asymetrically. This competition between the applied magnetic field and Coulomb correlations leads to interesting physics. The interband Coulomb interaction exponentially enhances the excitonic average. When the periodicity of the magnetic flux is commensurate with the lattice, the excitonic condensation is homogeneous throughout real space, whereas a one-dimensional modulation of order parameter is observed for incommensurate flux periodicity. The orbital magnetic field has a localizing effect on mobile $d$-electrons, affecting the excitonic coherence; a drop in the value of excitonic order parameter with both commensurate and incommensurate flux results. With non-zero $f$-electron hopping, further reduction in the excitonic averages is observed. The orbital field opposes the formation of an exciton as the kinetic energy of participating particles becomes larger than the condensation energy. In the small hybridization limit however, there is a different role of orbital field favouring the formation of an exciton at a special value $\alpha = 1/3$ per plaquette, thereby opening the possibility of tuning ferro-electricity via magnetic field.
\\
\section{Acknowledgements}
 SP acknowledges Center for Theoretical Studies, IIT Kharagpur for computer facilities. We acknowledge useful discussions with  Monodeep Chakraborty.

\end{document}